\documentstyle[preprint,aps]{revtex}

\newcommand{\str}        {$\rm^{1}$}
\newcommand{\legnaro}    {$\rm^{2}$}
\newcommand{\clt}        {$\rm^{3}$}
\newcommand{\bucarest}   {$\rm^{4}$}
\newcommand{\zagreb}     {$\rm^{5}$}
\newcommand{\itep}       {$\rm^{6}$}
\newcommand{\gsi}        {$\rm^{7}$}
\newcommand{\dresde}     {$\rm^{8}$}
\newcommand{\budapest}   {$\rm^{9}$}
\newcommand{\warsaw}     {$\rm^{10}$}
\newcommand{\rrik}       {$\rm^{11}$}
\newcommand{\heidelberg} {$\rm^{12}$}
\newcommand{\nantes}     {$\rm^{13}$}

\begin{document}
\draft

\title{Azimuthal Anisotropies as Stringent Test for Nuclear Transport Models}

\author{
P.~Crochet\str\footnote{Present address : GSI, Darmstadt, Germany},
F.~Rami\str,
R.~Don\`a\str$^{,}$\legnaro,
J.P.~Coffin\str,
P.~Fintz\str,
G.~Guillaume\str,
F.~Jundt\str,
C.~Kuhn\str,
C.~Roy\str,
B.~de~Schauenburg\str,
L.~Tizniti\str,
P.~Wagner\str,
J.P.~Alard\clt,
A.~Andronic\bucarest,
Z.~Basrak\zagreb,
N.~Bastid\clt,
I.~Belyaev\itep,
A.~Bendarag\clt,
G.~Berek\budapest,
D.~Best\gsi,
J.~Biegansky\dresde,
A.~Buta\bucarest,
R.~\v{C}aplar\zagreb,
N.~Cindro\zagreb,
P.~Dupieux\clt,
M.~D\v{z}elalija\zagreb,
Z.G.~Fan\clt,
Z.~Fodor\budapest,
L.~Fraysse\clt,
R.P.~Freifelder\gsi,
A.~Gobbi\gsi,
N.~Herrmann\gsi,
K.D.~Hildenbrand\gsi,
B.~Hong\gsi,
S.C.~Jeong\gsi,
J.~Kecskemeti\budapest,
M.~Kirejczyk\warsaw,
P.~Koncz\budapest,
M.~Korolija\zagreb,
R.~Kotte\dresde,
A.~Lebedev\itep$^{,}$\rrik,
Y.~Leifels\gsi,
V.~Manko\rrik,
D.~Moisa\bucarest,
J.~M\"osner\dresde,
W.~Neubert\dresde,
D.~Pelte\heidelberg,
M.~Petrovici\bucarest,
C.~Pinkenburg\gsi,
W.~Reisdorf\gsi,
J.L.~Ritman\gsi,
A.G.~Sadchikov\rrik,
D.~Sch\"ull\gsi,
Z.~Seres\budapest,
B.~Sikora\warsaw,
V.~Simion\bucarest,
K.~Siwek-Wilczy\'nska\warsaw,
U.~Sodan\gsi,
K.M.~Teh\gsi,
M.~Trzaska\heidelberg,
G.S.~Wang\gsi,
J.P.~Wessels\gsi,
T.~Wienold\gsi,
K.~Wisniewski\gsi,
D.~Wohlfarth\dresde,
A.~Zhilin\itep \\
(FOPI Collaboration) \\ 
and C.~Hartnack\nantes\\
}

\address{
\str  Institut de Recherches Subatomiques, IN2P3-CNRS/ULP, Strasbourg, France \\
\legnaro Istituto Nazionale di Fisica Nucleare, Legnaro, Italy \\
\clt Laboratoire de Physique Corpusculaire, IN2P3-CNRS,
Universit\'e Blaise Pascal, Clermont-Ferrand, France\\
\bucarest Institute for Physics and Nuclear Engineering, Bucharest, Romania \\
\zagreb Ru{d\llap{\raise 1.22ex\hbox
{\vrule height 0.09ex width 0.2em}}\rlap{\raise 1.22ex\hbox
{\vrule height 0.09ex width 0.06em}}}er
Bo\v{s}kovi\'{c} Institute, Zagreb, Croatia \\
\itep Institute for Theoretical and Experimental Physics, Moscow, Russia \\
\gsi Gesellschaft f\"ur Schwerionenforschung, Darmstadt, Germany \\
\dresde Forschungszentrum Rossendorf, Dresden, Germany \\
\budapest Research Institute for Particles and Nuclear Physics,
Budapest, Hungary \\
\warsaw Institute of Experimental Physics, Warsaw University, Warsaw, Poland \\
\rrik Russian Research Institute ``Kurchatov", Moscow, Russia \\
\heidelberg Physikalisches Institut der Universit\"at Heidelberg,
Heidelberg, Germany \\
\nantes SUBATECH, UMR Universit\'e, IN2P3-CNRS, Ecole des Mines,
Nantes, France \\
}

\maketitle

\begin{abstract}
Azimuthal distributions of charged particles and intermediate mass
fragments emitted in  Au+Au collisions at 600$A$\,MeV
have been measured using the FOPI facility at GSI-Darmstadt.
Data show a strong increase of the in-plane azimuthal anisotropy ratio with the
charge of the detected fragment. Intermediate mass fragments are found
to exhibit a strong momentum-space alignment with respect of the reaction
plane.
The experimental results are presented as a function of 
the polar center-of-mass angle and over a broad range of impact 
parameters.
They are compared to the predictions of the Isospin Quantum
Molecular Dynamics model using three different
parametrisations of the equation of state.
We show that such highly accurate data provide
stringent test for microscopic transport models and can potentially
constrain separately the stiffness of the nuclear equation of state
and the momentum dependence of the nuclear interaction.
\end{abstract}

\vspace{0.5cm}


\vspace{1.cm}


\section{Introduction}

One of the major objectives in the investigation of high energy
heavy ion collisions ($E_{\rm lab} >$ 100$A$\,MeV) 
is to provide information on the
equation of state (EoS) of nuclear matter
under extreme conditions of density and temperature.
In this respect, collective flow effects of various
ejectiles emerging from the reaction are of great interest since they are
known as a sensitive probe of compression of nuclear matter~\cite{sto 80}.
These collective phenomena were predicted by nuclear hydrodynamical 
models~\cite{sto 80,sch 74} and experimentally identified in the eighties 
with the advent of the first generation of $4\pi$ detectors 
(Plastic Ball~\cite{bad 82} and Streamer Chamber~\cite{sch 79} 
at LBL-BEVALAC and
the DIOGENE detector~\cite{ala 87} at SATURNE) capable of full event by event
characterisation. Different forms of nuclear matter flow were observed :
i) the directed sideward flow which appears as a sideward deflection 
of the emitted particles in the reaction plane~\cite{gus 84,ren 84} 
ii) the squeeze-out
effect which manifests itself as an enhanced emission out of the reaction 
plane~\cite{gut 89b,dem 90} and iii) the so-called radial flow which 
has been more recently evidenced by the FOPI collaboration as an azimuthally
symmetric expansion in highly central collisions~\cite{jeo 94}. 

Despite the extensive experimental studies~\cite{rev 11}
and the substantial theoretical
progress~\cite{aic 87,pei 89,zha 94,dan 95,pan 93,aic 91,cas 90,fuc 96,wel 88} 
achieved so far, an unambiguous determination of the nuclear EoS
from heavy ion collision studies 
is not yet at hand to date.
Indeed there remain still several difficulties in addressing this 
intriguing question :

\begin{itemize}

\item The extraction of information from experimental observables
demands a thorough understanding of the reaction dynamics which can be
possible only through prerequisite tests of nuclear 
transport models.

\item{The experimentally reconstructable observables are not uniquely 
sensitive to the incompressibility coefficient ($K$) characterising 
the nuclear EoS. Other effects
such as the momentum dependence of the nuclear interaction (MDI) and 
the in-medium modification of the nucleon-nucleon cross section 
($\sigma_{\rm nn}$) play
also an important role~\cite{aic 87,pei 89,aic 91,gal 87}.
It is therefore
crucial to disentangle the influence of the EoS from the one related
to the other physical effects above mentioned.} 

\item{There are also several input parameters needed for the initialisation
of transport codes~\cite{har 96} and somewhat arbitrarly fixed in the 
calculations, which might also influence significantly the theoretical
predictions.}   

\item{The sensitivities of the experimental flow observables to the EoS
(as well as MDI and 
in-medium ${\sigma}_{\rm nn}$) 
are generally quite 
low~\cite{pei 89,ramil 95,wie 93},
which requires highly accurate data. This implies in particular a
good accuracy on the reaction plane reconstruction and low detector
biases and inefficiences.}    

\item{Available transport models underpredict composite particles 
yields~\cite{tsa 93,rei 96,gos 95,ram 96}. 
Thus, a meaningful comparison between experimental flow 
observables and those predicted by the theory should be made on the basis of 
coalescence invariant quantities including the contributions of all
emitted particles. 
This requires in particular the inclusion of intermediate mass fragments 
(IMF, $Z\ge 3$)
which are known to carry larger amount of flow than do lighter 
particles~\cite{dos 87,par 95,hua 96,ram 93}
and which constitutes a non negligeable fraction of the total mass of the
system even at incident energies of a few hundred 
$A$\,MeV~\cite{rei 96,ram 93,ala 92,her 93}.}

\end{itemize}

Several experimental observables, corresponding to different projections
of the triple differential cross section, have been introduced in order
to characterise quantitatively the strength of the collective flow in 
heavy ion collisions~\cite{gut 89b,dan 85,gyu 82}. 
One of these observables is the so-called ``in-plane azimuthal anisotropy 
ratio" (defined in section IV)~\cite{wel 88} which can be obtained from 
measurements of azimuthal distributions relative to the reaction plane.
Theoretical calculations~\cite{wel 88,schu 87} in the framework of the 
Boltzmann-Uehling-Uhlenbeck (BUU) model~\cite{ber 88}
have shown the relevance of this observable as a sensitive quantity
to pin down the nuclear EoS. 
Comparisons between data and BUU 
calculations were carried out for neutrons~\cite{mad 93,ela 94}, 
but in this case 
the sensitivity to variations 
of the equation of state was found to be rather low.
Recently, it was reported~\cite{zha 95} that IMF, because of their strong 
alignment along the flow direction~\cite{dos 87}, 
correlate much better with the value of $K$ used in the theory.  

We report in this paper the first detailed experimental results 
on the in-plane azimuthal anisotropy ratio of light charged particles and IMF.
The data were obtained from azimuthal distribution 
measurements carried out for the Au+Au reaction at several bombarding
energies from 100 to 800$A$\,MeV, using the FOPI facility at the SIS 
(GSI-Darmstadt) accelerator. Here we focus 
on the high incident energy of 600$A$\,MeV where : i) the 
participant and the spectator products are better 
separated, ii) compressional and MDI effects are expected to be 
large~\cite{aic 87,aic 91} and iii) high statistics dynamical model 
calculations are available~\cite{har 92b}.
We find that the in-plane anisotropy ratio increases with the charge of the
detected fragment. We show also that IMF exhibit a very pronounced
alignment along the flow direction. 
The dependencies of the anisotropy ratio on the polar angle 
in the center-of-mass (c.m.) frame 
and the collision centrality are presented. They are compared to
the predictions of the Isospin Quantum Molecular Dynamics 
(IQMD)~\cite{har 92} model for different parametrisations of 
the equation of state. Simulated events are filtered
through the FOPI acceptance and sorted in accordance to the adopted
experimental procedure, which allows direct and realistic comparisons 
between data and model predictions.
We show that the availability of such experimental data provide a stringent 
test for dynamical transport models.      
Based on FOPI data it has been shown that the shape of the fire-ball
in central collisions depends sensitively on $\sigma_{\rm nn}$\cite{roy 97}.
Here, we emphasize the possibility to disentangle compressional and MDI 
effects on the basis of detailed azimuthal distributions in non central 
collisions for the participant as well as for the spectator source.

\section{Experimental setup}

The data were obtained in a 
complete experimental study of the Au+Au system at 6 beam energies 
going from 100 to 800$A$\,MeV at the SIS (GSI-Darmstadt) accelerator using 
the FOPI/phase-I facility. As already noted, the present work is 
restricted to the incident energy of 600$A$\,MeV. 
At this energy, about ${10}^6$ events
were collected under the central trigger condition ~\cite{gob 93}. 
The latter was defined by adjusting the charged particle multiplicity
to a value which corresponds to impact parameters less than $9 fm$ in 
a clean-cut geometrical model.   
Most of the data which will be presented in the following
were obtained from the analysis of these central events. Only those 
related to the centrality dependence will include events taken under the
minimum bias condition~\cite{gob 93} ({\em i.e.} events 
where at least 4 charged 
particles were detected in the angular range 
$7^{\circ}\le \Theta_{\rm lab} \le 30^{\circ}$). 
This large amount of available events allowed us to extract high statistics
data.    

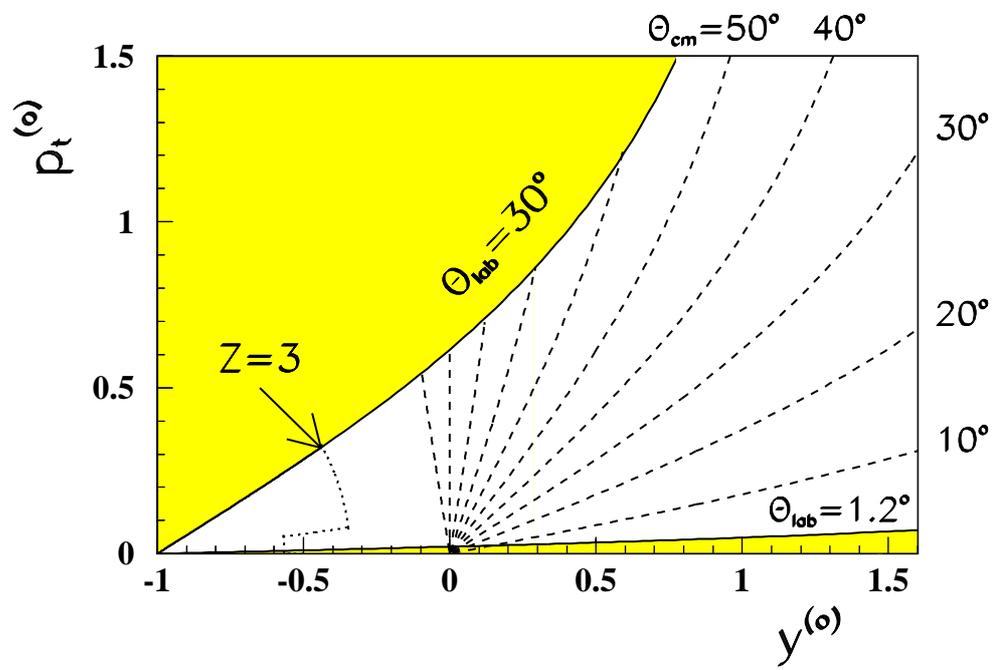
\begin{figure}
\caption{Illustration of the
FOPI/Phase-I acceptance in the ($p_{\rm t}^{(0)}, y^{(0)}$)
plane for Au+Au collisions at 600$A$\,MeV (see text for more details).}
\label{fig1}
\end{figure}

The FOPI detector has been described in details in reference~\cite{gob 93}.
Here, we recall briefly its different components and main features. 
At the incident energy considered in this paper, the setup consisted 
of a highly segmented Forward Wall of plastic scintillators, covering in full 
azimuth the laboratory polar angles ($\Theta_{\rm lab}$) 
from $1.2^{\circ}$ to $30^{\circ}$,
complemented by 
a shell of thin plastic scintillator paddles mounted in front of it
and subtending the
$\Theta_{\rm lab}=1.2^{\circ}$ - $7^{\circ}$ angular range.
This device allowed us, event by event, to identify the nuclear charge 
and measure the vector velocities of most of the light charged
particles and IMF (up to $Z=12$)
emitted in the forward c.m. hemisphere.
Its high granularity (about a factor of three higher than the one of the
earlier Plastic Ball detector~\cite{bad 82}) allowed high multiplicity events
to be measured with a low multi-hit rate. 
The apparatus provides detailed and highly accurate measurements of triple
differential cross sections for all charged particles with a very good 
accuracy on the reaction plane determination which is of crucial importance
in the present work.   
Figure~\ref{fig1} illustrates the acceptance of the Phase-I 
configuration of the FOPI 
detector in the 
($p_{\rm t}^{(0)},y^{(0)}$) plane, where
$p_{\rm t}^{(0)}$  is the transverse momentum per nucleon
normalized to the c.m. projectile momentum per nucleon ($=528.6$\,MeV/c)
and $y^{(0)}$ denotes the rapidity normalized to the projectile
rapidity in the c.m. frame.
Solid curves represent the polar geometrical limits of the FOPI apparatus
in the laboratory frame ($\Theta_{\rm lab}=1.2^{\circ}$ and $30^{\circ}$).
The detection thresholds are indicated by the dotted curves in the case of 
$Z=3$ fragments. The dashed curves mark different c.m. polar 
angles ($\Theta_{\rm c.m.}$) separated by 10 degrees.
As it can be seen in Fig.~\ref{fig1}, the apparatus allows undistorded 
measurements (free of detector biases) to be performed over a broad region 
of the forward c.m. hemisphere.

\section{Event characterisation}

 The study of azimuthal distributions requires a good event 
characterisation both in impact parameter and azimuthal orientation 
of the reaction plane. This is also of crucial importance when comparing
the experimental data to theoretical predictions.
In this respect, it is interesting to point out 
that a heavy system such as Au+Au has the advantage to offer
a better determination of the collision geometry, both in centrality 
and reaction plane orientation, than lighter systems.
In this section we will describe the procedure used in  
in order to characterize the measured events and illustrate the
performances of the FOPI apparatus. 

\vspace{0.5cm}

To sort out the events according to their degree of centrality, 
we have employed the standard method based on the correlation between 
the multiplicity of the emitted particles and the impact parameter. 
The multiplicity distribution of charged particles measured 
in the angular range $7^{\circ}\le \Theta_{\rm lab} \le 30^{\circ}$ 
exhibits a typical plateau for intermediate values and falls off for
the highest multiplicities~\cite{rei 96,ala 92}.
This distribution was divided into five intervals, 
in accordance to the procedure introduced in previous 
works~\cite{ala 92,dos 85}.  
The highest multiplicity bin (named PM5) starts at half the plateau value
corresponding to 
$\simeq 2.5\%$ of the total reaction cross section and 
a maximum impact parameter of $\simeq 2.5fm$ in a sharp cut-off 
approximation. 
The remaining multiplicity range was divided into four equally
spaced intervals named PM1 to PM4. The limits of these event classes
are given in 
Tab.~\ref{tab1}. 
In the following only events belonging to the PM2-PM5 bins
are considered. 
An additional condition requiring the total detected charge to be larger
than 30 was imposed in the analysis, in order to reduce the contribution
of the
background contamination from non target collisions.
Within this condition the background contamination 
was estimated, by comparing measurements 
with and without target,  
to be less than $5\%$ in the PM2 multiplicity range and  
negligible for higher multiplicity events.
The mean impact parameters and the associated r.m.s. deviations 
listed in Tab.~\ref{tab1} 
were extracted from IQMD simulations, where the theoretical events were
passed through the detector acceptance
and sorted in accordance to the above outlined
procedure used for the data.  
By doing so, one can compare directly the experimental observables to the 
the theoretical ones, 
even if the calculations do not reproduce the measured
multiplicity distribution, which is the case of the IQMD model.
The latter, as we will see in section V, underpredicts IMF multiplicities
and overpredicts the overall charged particle multiplicities.
This can be seen in Tab.~\ref{tab1}, where the lower limit of the 
PM5 bin for IQMD events is much larger (by 14 units) than the 
experimental one.  
However, when scaled to the lower limit of the PM5 bin, the theoretical 
multiplicity distribution is found to be in a fairly good agreement with 
the data in the range PM2-PM5. 
As shown in Tab.~\ref{tab1}, the charged particle multiplicity
criterium covers a rather wide impact parameter 
range from $<b>\simeq 3 fm$ to $<b>\simeq 9 fm$.

\vspace{0.5cm}

The reconstruction of the azimuth of the reaction plane have been 
performed according to the 
transverse momentum analysis~\cite{dan 85}. 
In the framework of this method, the reaction plane is determined,
for each particle $\mu$ in a given event, by
the vector $\vec{Q}$ calculated from the transverse 
momenta $\vec{p_{\rm t}^{\nu}}$ of all detected particles 
except the particle $\mu$ in order to remove autocorrelation effects :
$$\vec{Q}\;=\;\sum_{{\rm \nu=1\atop \nu\neq \mu}}^{\rm M}\,\omega_{\nu} 
\vec{p_{\rm t}^{\nu}}$$
M is the multiplicity of the event  
and $\omega_{\nu} = 1\;\mbox{if}\;y^{(0)}>\delta, 
-1\;\mbox{if}\;y^{(0)}<\delta$ and $0$ otherwise.
$\delta$, choosen equal to 0.5 in the present work, 
is a parameter introduced to remove mid-rapidity particles which are 
less correlated with the reaction plane.
In order to estimate the accuracy on the reaction plane determination,
due to finite number of particle effects and detector biases, 
we have used the method described in reference~\cite{dan 85} which
consists in randomly subdividing each event into two and calculating on average
the half difference in azimuth ($\Delta {\Phi}_{\rm R}$) 
between the reaction planes 
extracted from the two sub-events. 
$\Delta {\Phi}_{\rm R}$ gives an estimate of the
dispersion of the reconstructed reaction plane with respect 
to the true one~\cite{dan 85}.
The results are displayed in Tab.~\ref{tab1} in terms of the 
standard deviation 
width $\sigma(\Delta {\Phi}_{\rm R})$ 
extracted from a gaussian fit to the 
$\Delta\Phi_{\rm R}$ distributions. 
As can be seen, the reaction plane is, in all cases, rather well 
estimated, with a precision which varies typically from $\simeq 24^{\circ}$
to $\simeq 34^{\circ}$ depending on the event multiplicity.
The data presented in this paper are corrected for these uncertainties
on the reaction plane determination (see below). 

\section{Experimental results}

Azimuthal distributions are generally used to study the emission pattern of
particles in the participant region ({\em i.e.} at mid-rapidity) of the 
collision.
Instead of restricting the analysis to a narrow rapidity window 
centered around the c.m. rapidity, we have explored here
the whole experimentally covered phase space (see Fig.~\ref{fig1}). 
To select particles emitted in different regions of the phase space
the analysis was carried out by imposing c.m. polar angle 
($\Theta_{\rm c.m.}$) gates to the data, as suggested in 
previous works~\cite{tsa 93,ela 94}.
The use of this $\Theta_{\rm c.m.}$ binning 
was found to be more suited than using 
rapidity cuts because of the typical FOPI/phase-I acceptance 
(Fig.~\ref{fig1})
which is limited to laboratory polar angles lower than 
$30^{\circ}$.
Indeed by doing so, one can extract an accurate quantitative information
from the measured triple differential cross sections over a broad region 
of the forward c.m. hemisphere 
where, as we will see in the following, 
the data are only very weakly affected by the detector cuts.  

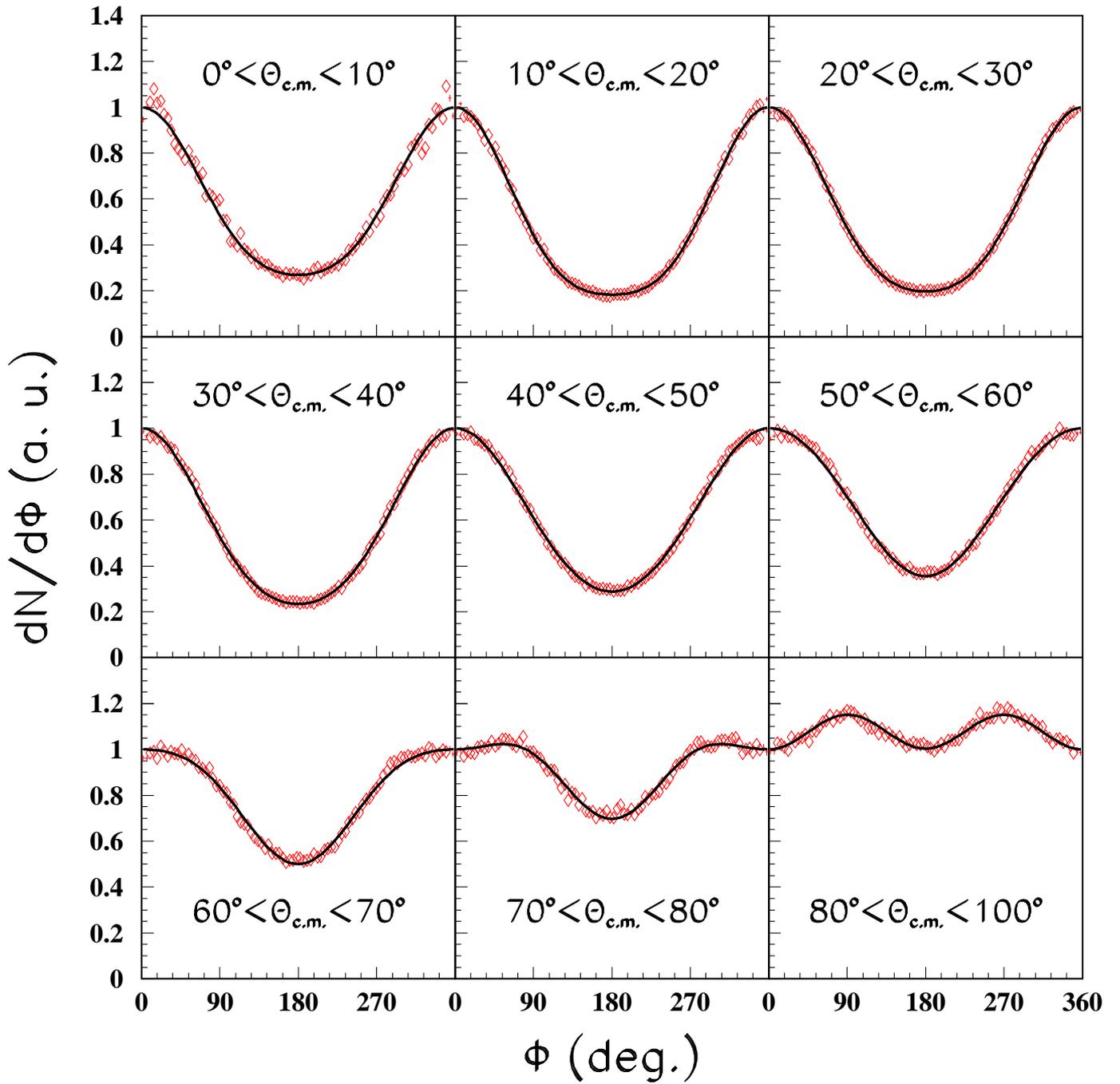
\begin{figure}
\caption{Azimuthal distributions, with respect of the reaction
plane, including the contribution of all detected charged
particles, each being weighted by its charge. The data are shown
for the semi-central PM4 event class, for different $\Theta_{\rm c.m.}$ bins
reported on the plots.
Data points are larger than the corresponding statistical uncertainties.
The curves are the results of the fit described in the text.}
\label{fig2}
\end{figure}

Typical examples of azimuthal distributions around the beam axis 
$Y(\Phi) = dN/d\Phi$
(where $\Phi$ is the azimuth of the emitted particle 
with respect to the azimuthal orientation of the reaction plane) 
of charged particles 
are displayed in Fig.~\ref{fig2} for the PM4 event class. 
These distributions were obtained for proton-like particles, {\em i.e.} by
including all detected particles each being weighted with its charge.
Different panels in Fig.~\ref{fig2} show results for different c.m. 
polar angle gates each 10 degrees wide except for the one around 
${90}^{\circ}$ which was taken from ${80}^{\circ}$ to ${100}^{\circ}$. 
With increasing $\Theta_{\rm c.m.}$, a drastic change in the azimuthal emission 
pattern is observed.
At forward polar angles, the distribution exhibits a strongly enhanced
in-plane emission along the sideward flow direction
($\Phi=0^{\circ}$),
while at large polar $\Theta_{\rm c.m.}$ angles (approaching 
$\Theta_{\rm c.m.}=90^{\circ}$), one observes a clear preferential emission
in the out-of-plane ($\Phi=90^{\circ}$ and $\Phi=270^{\circ}$) 
direction corresponding to the squeeze-out effect.
Thus, the shape of these azimuthal distributions reflects the strength 
of the collective motion for both the 
directed in-plane flow and the out-of-plane
emission. In a previous paper~\cite{bas 96} we have reported the 
results concerning the preferential out-of-plane emission. 
Here we focus on the in-plane anisotropy component. 

In order to examine the data in a quantitative manner, we have 
extracted the magnitude of the in-plane anisotropy $R$ for each 
$\Theta_{\rm c.m.}$
bin as :
\begin{equation}
R\;=\;\frac{\left. Y(\Phi) \,\right|_{0^{\circ} < 
\Phi < 45^{\circ}} +
\left. Y(\Phi) \,\right|_{315^{\circ} < \Phi < 360^{\circ}}}
{\left. Y(\Phi) \,\right|_{135^{\circ} < \Phi < 225^{\circ}}}
\end{equation}  
This definition of the in-plane anisotropy signal differs from the 
one used in earlier works~\cite{wel 88,ela 94}, namely the ratio
$\left. Y(\Phi) \,\right|_{\Phi=0^{\circ}} \,/\,
\left. Y(\Phi) \,\right|_{\Phi=180^{\circ}}$. 
This was dictated by statistics limitations for
the theoretical IQMD calculations. 

\vspace{0.5cm}

Fluctuations of the reconstructed reaction planes with respect to the true 
ones cause an attenuation of the experimentally observed anisotropies.
To take into account this effect, we have proceeded in the following 
way~\cite{cro 96} :

\begin{itemize}
\item {We have first fitted a function of the form 
$a_{\rm 0}\,+\,a_{\rm 1}\cos(\Phi)\,+\,a_{\rm 2}\cos(2 \Phi)$
to the measured $Y(\Phi)$ distributions. The resulting fits 
are shown by the curves in Fig.~\ref{fig2}.} 
\item {Then following~\cite{dem 90}, 
we have determined the corrected parameters : 

$a^{\prime}_{\rm 0}\;=\;a_{\rm 0}\quad,\quad
a^{\prime}_{\rm 1}\;=\;a_{\rm 1}\,/\,\left(<\cos(\Delta{{\Phi}_{\rm R}})>\right)\quad,\quad
a^{\prime}_{\rm 2}\;=\;a_{\rm 2}\,/\,\left(2\,<\cos^2(\Delta{{\Phi}_{\rm R}})>
- 1\right)$.}
\item {Finally, we have extracted the corrected anisotropy ratios 
R by integrating in equation (1) the corrected fitting 
function.}    
\end{itemize}

The values of $\cos(\Delta{{\Phi}_{\rm R}})$ and 
$\cos^2(\Delta{{\Phi}_{\rm R}})$ were determined event wise and then
averaged for each event class.
The results are shown in Tab.~\ref{tab2} in the case of the 
$40^{\circ} < \Theta_{\rm c.m.} < 50^{\circ}$ gate
for different centrality bins.
As can be seen, the corrections become more important for peripheral 
events (PM2 bin)  
because of the relatively large dispersion in $\Delta\Phi_{\rm R}$. 

\vspace{0.5cm}

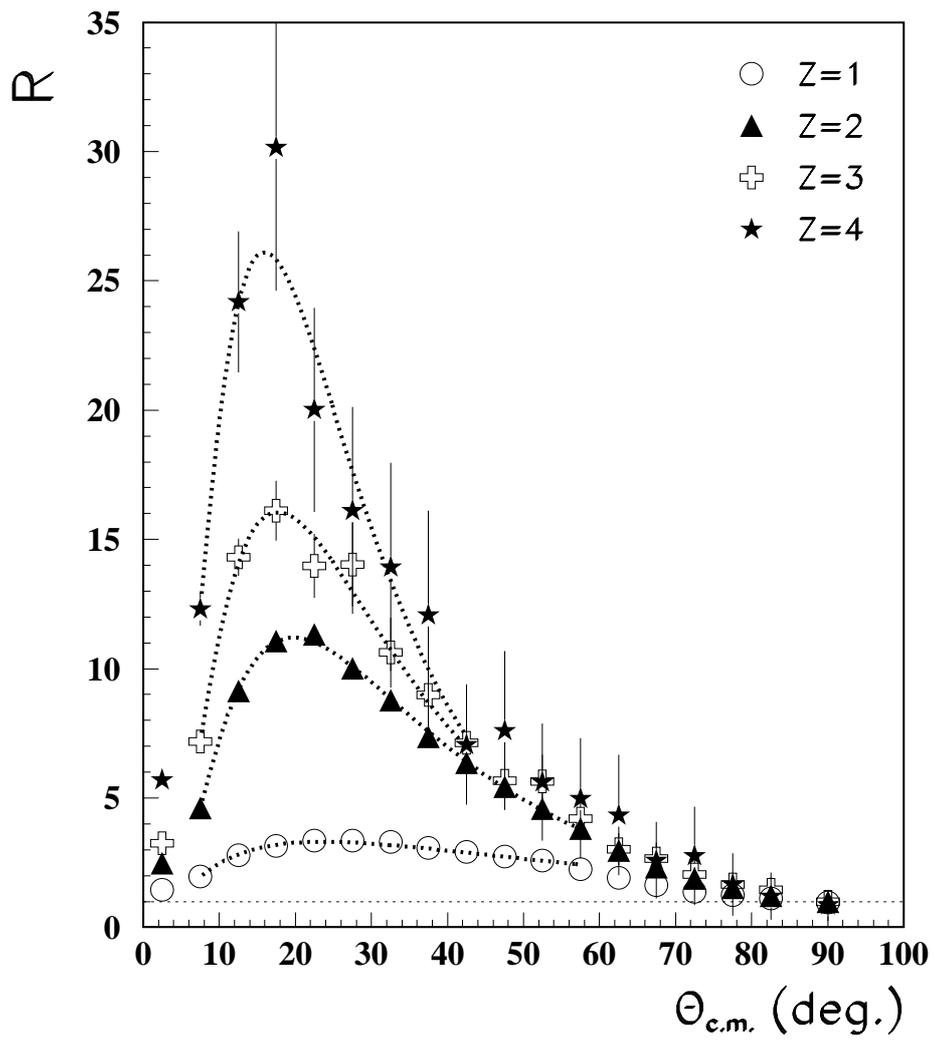
\begin{figure}
\caption{In-plane anisotropy ratio $R$ as a function of
the polar angle $\Theta_{\rm c.m.}$, for
different types of ejectiles. The data are shown for the PM4 event class.
The error bars are of statistical origin.
The dotted curves represents a gaussian fit to the correlation
between $R$ and $log_{\rm 10}(\Theta_{\rm c.m.})$.}
\label{fig3}
\end{figure}

Figure~\ref{fig3} shows the in-plane anisotropy ratio $R$ extracted 
from the data according to the definition in equation (1) and corrected
for fluctuations of the reconstructed reaction plane, as a function of
the polar angle $\Theta_{\rm c.m.}$. 
The results were preselected by the charged particle multiplicity cut PM4.
They are presented for different types of particles ($Z=1$ to 4).   
The observed trend is characterized by a well defined maximum located
at forward c.m. polar angles ($\Theta_{\rm c.m.} \simeq 15^{\circ}$). 
The presence of this maximum is related
to the bounce-off of the projectile remnants. 
For polar angles approaching $\Theta_{\rm c.m.}=90^{\circ}$, the anisotropy 
ratio tends to $R=$1 as expected
from symmetry reasons. 
Here, the azimuthal distribution
exhibits a completely different pattern with an enhanced out-of-plane
emission (Fig.~\ref{fig2}) whose magnitude is commonly 
characterized by the squeeze-out ratio~\cite{dem 90,gut 90}.
It should be pointed out that the overall trend, observed in Fig.~\ref{fig3}, 
is qualitatively consistent with earlier data~\cite{tsa 93a}
obtained at much lower 
beam energy.

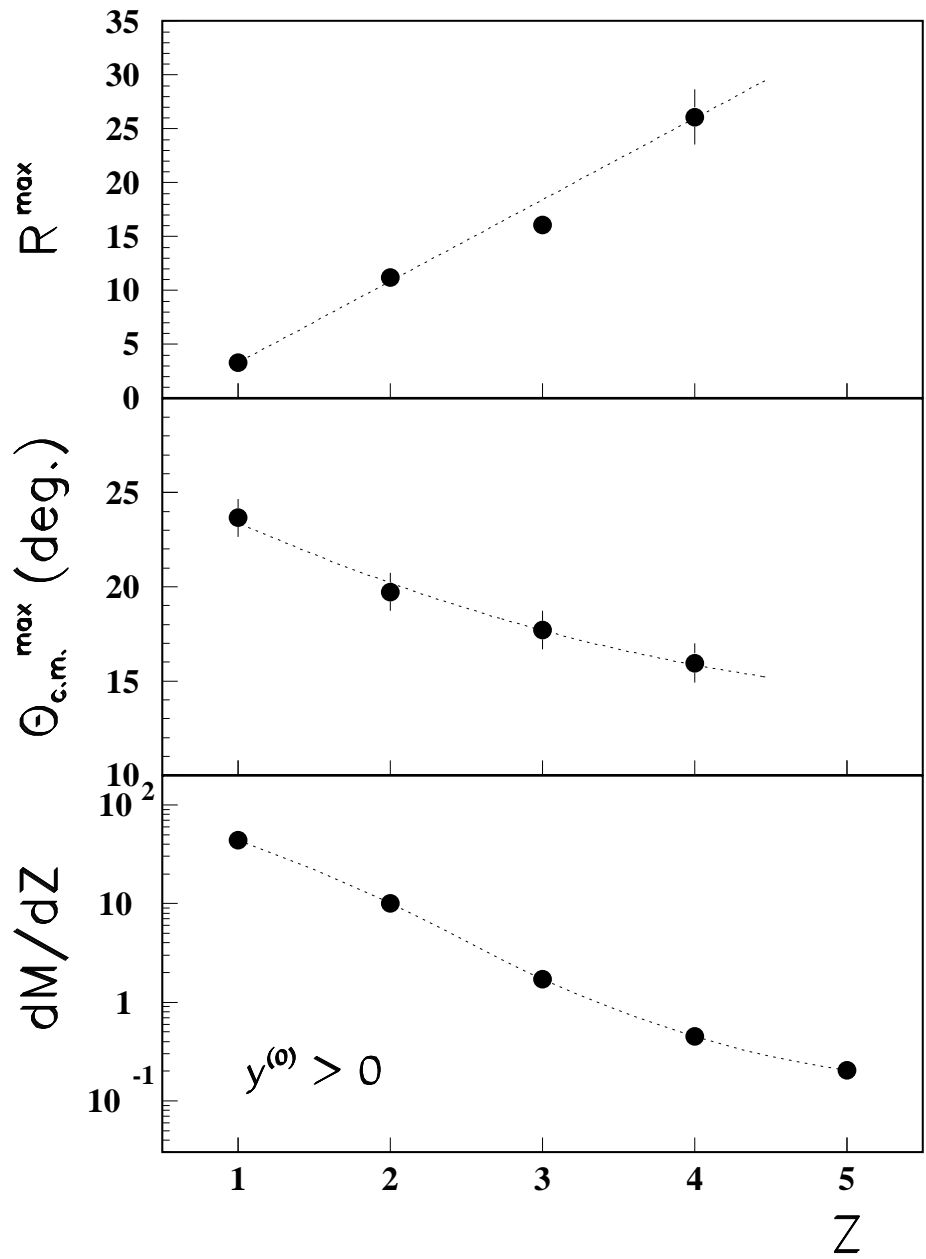
\begin{figure}
\caption{Maximum in-plane anisotropy $R^{\rm max}$ (top-panel) and 
the corresponding polar
angle $\Theta_{\rm c.m.}^{\rm max}$ (mid-panel) as a function of
the charge of the detected fragment. The lower panel shows the $Z$ dependence
of the mean multiplicity in the forward c.m. hemisphere.
The data are shown for the PM4 event class.
Error bars include statistical errors only.
The curves are just to guide the eye. In the lower panel, data points are
larger than the corresponding statistical uncertainties.}
\label{fig4}
\end{figure}

As can be seen in Fig.~\ref{fig3}, with increasing fragment charge one observes 
larger azimuthal asymmetries with a more pronounced maximum whose location
is shifted towards smaller polar angles. 
These features are more quantitatively illustrated in Fig.~\ref{fig4}, where
the maximum $R^{\rm max}$ (top-panel) 
of the correlation 
$R$ versus $\Theta_{\rm c.m.}$ and its location $\Theta_{\rm c.m.}^{\rm max}$ 
(mid-panel) 
are plotted as a function of the
charge of the detected fragment. $R^{\rm max}$ and $\Theta_{\rm c.m.}^{\rm max}$
were extracted by fitting a gaussian to the measured 
correlation between $R$ and ${log}_{\rm 10}(\Theta_{\rm c.m.})$. 
The resulting fits are 
shown, in the angular range where the fitting procedure was applied, 
by the dotted curves in Fig.~\ref{fig3}.
We have also displayed in Fig.~\ref{fig4} (lower-panel) the mean 
fragment multiplicities 
in the forward c.m. hemisphere ($y^{(0)} > 0$). 
It is worth to note that IMF are still 
observed with quite large multiplicities at the relatively high 
incident energy of 600$A$\,MeV explored in the present work. 
We found~\cite{don 94} that IMF carry about $12\%$ of the total detected 
charge in the case of the PM4 event class.
It should be noted that for this class
of events, a substantial fraction of IMF originates from the decay of the
spectator source. When the latter source is selected
the fraction of IMF remains nearly the same in the whole beam energy range
from 250 to 800AMeV~\cite{bea 97}, what is in agreement with the universal 
behaviour of the spectator decay reported by the ALADIN 
collaboration~\cite{alad 96}.
The observed dependence between $R^{\rm max}$ and $Z$ (top-panel in 
Fig.~\ref{fig4})  
 shows a roughly linear rise of the maximum anisotropy ratio with 
increasing fragment charge at least up to $Z=4$. 
Above $Z=4$, statistics in the 
data was not sufficient to extract accurate $R$ values, because of 
the very large anisotropies and the low multiplicities. 
On the other hand, the mid-panel of Fig.~\ref{fig4} shows that with increasing
$Z$ the location $\Theta_{\rm c.m.}^{\rm max}$
of the maximum anisotropy moves towards lower values 
and seems to saturate. 

The observed dependencies of 
$R^{\rm max}$ and $\Theta_{\rm c.m.}^{\rm max}$ with the fragment charge, 
reflect the strong alignment of IMF along the flow 
direction, an effect which was qualitatively observed in the earlier
Plastic Ball experiments~\cite{dos 87}  for the same Au+Au system at 
a beam energy of 200$A$\,MeV. 
They are consistent with the increase of the sideward flow with 
the fragment size which was also reported in reference~\cite{dos 87} and 
recently confirmed in more quantitative studies by 
several groups~\cite{par 95,hua 96,ram 93,don 94,don 96} .
The strong alignment of IMF could be qualitatively
understood~\cite{dos 87,cro 96} from a simple picture where the
pure collective motion is superimposed upon the random thermal one :
light particles being more sensitive to thermal fluctuations which tend to
wash out their flow,
while IMF being less affected by the undirected thermal motion are expected
to be much more aligned along the flow direction.

\section{Comparison to IQMD model predictions}

In this section we compare the experimental results to the 
predictions of the QMD model~\cite{pei 89,aic 91,har 92,sto 86}
using the so-called IQMD version~\cite{har 92}.
Besides the test of the IQMD model, our main aim here is to examine
the possibility to disentangle compressional (EoS) and MDI effects on 
the basis of the present experimental results. 
It is worth to recall (see section I) that the theoretical predictions might 
also depend sensitively on other ingredients in the model such as
the in-medium nucleon-nucleon cross section
($\sigma_{\rm nn}$) which is not investigated in the present work.
The possibility to constrain the in-medium $\sigma_{\rm nn}$ has been 
discussed in previous works~\cite{pei 89,roy 97}. 
We start with a very short description of the model and recall some
of the typical features in the IQMD version. Then we discuss the 
influences of the detector biases on the extraction of the in-plane 
anisotropy ratio. Finally,  
we confront the data to the model calculations
for different parametrisations of the nuclear EoS.       

\subsection{The IQMD model}

The Quantum Molecular Dynamics (QMD) 
model~\cite{pei 89,aic 91,har 92,sto 86} is a n-body theory which calculates
the time evolution of intermediate energy heavy ion reactions on an
event by event basis.
Nucleons are represented as gaussian wave functions whose centroids
are propagated according to the classical equations of motion and
their width parameter $L$ is kept fixed.
These wave packets are submitted to two and three body local Skyrme forces
plus Yukawa and long range Coulomb potentials.
An additional term describing the momentum dependence of the nuclear interaction
(MDI) is included.
The latter, which is computed from the momentum dependence of the optical
p-nucleus potential, produces an additional transverse deflection of nucleons 
early in the reaction.
The parameters of the density dependent 
Skyrme and MDI potentials 
are adjusted to reproduce the ground state properties
of the infinite nuclear matter and to fix the choice of the incompressibility
$K$ of the EoS.
The numerical simulation of a complete event includes the initialisation
procedure of projectile and target (see below), the propagation of nucleons
through the previous potentials 
and the nucleon-nucleon interactions {\em via} a stochastic scattering term.
For each binary collision, the phase space distributions in the final stage
of the scattering partners are checked to obey the Pauli principle, 
otherwise the collision is blocked.
At the final stage of the reaction ($t=200fm/c$), 
composite particles are formed 
if the centroid distances are lower than $3 fm$.

The IQMD (Isospin Quantum Molecular Dynamics)~\cite{har 92}
is a QMD version which takes into account isospin degrees of freedom
for the cross sections and the Coulomb interaction.
This version uses a 
gaussian width fixed 
to $4L = 8.66fm^2$
for Au nuclei.
The initialisation procedure consists in distributing randomly
into a spherical phase space the position
$r$ and the momentum $p$ of the gaussian centroids 
with $r \le R$ and $p \le p_{\rm F}$, where $R=1.12A^{1/3}$ 
($A$ being the total 
nuclei mass number) and $p_{\rm F} = 268$MeV/c (Fermi momentum).

For the present work, the choice of this version, among other different
realisations of the QMD model, is justified by the fact that IQMD 
is particularly suited
for the analysis of nuclear matter collective 
effects~\cite{har 89,har 94a,har 94b,sof 95,bas 93,bas 94}.
Our study is however relevant in a more general level
since the major deviations about nuclear matter flow between the different
QMD versions are now known to stem exclusively 
from the initialisation procedures and the gaussian widths~\cite{har 96}.

\subsection{Details about the calculations}

IQMD events were computed~\cite{har 92b} for three different
parametrisations of the nuclear EoS : i) a hard EoS ($K=380$$A$\,MeV)
without MDI (H), ii) a hard EoS ($K=380$$A$\,MeV) with MDI (HM) and
iii) a soft EoS ($K=200$$A$\,MeV) with MDI (SM).
The calculations were performed for the full range of impact parameters
from 0 to 14$fm$.  
The number of simulated events was about 300-1000 per $fm$ interval.
It is important to mention that the theoretical events were triggered
using the same centrality selection as for the data.
In addition they were filtered through the acceptance of the FOPI/Phase-I
detector including geometrical cuts and energy thresholds.
The calculations were restricted to charged particles 
with $1 \le Z \le 12$.
Here, the azimuthal distributions were determined with respect to the true 
reaction plane (known in the model).

\vspace{0.5cm}

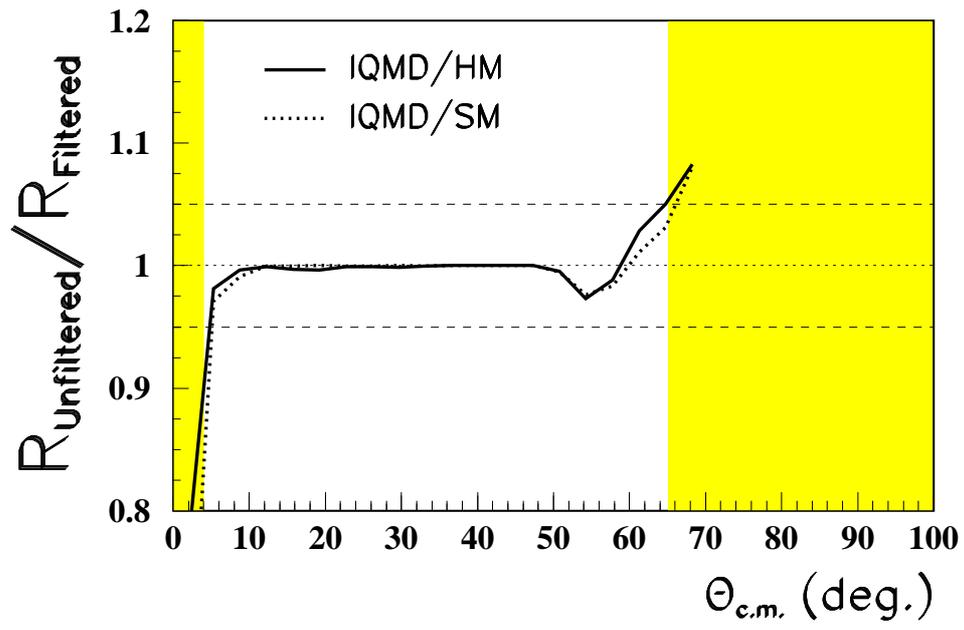
\begin{figure}
\caption{In-plane anisotropy $R$
obtained from unfiltered IQMD simulations divided by the one corresponding
to filtered calculations, as a function of the c.m. polar angle
$\Theta_{\rm c.m.}$.
The calculations are shown for Au(600$A$\,MeV)+Au
collisions preselected with the PM4 multiplicity cut.
The full (dotted) line corresponds to a HM (SM) parametrisation of the
nuclear EoS.
The horizontal dashed lines indicate $R = 1 \pm 0.05$ values.
Statistical uncertainties in the IQMD values are lower than $2\%$.}
\label{fig5}
\end{figure}

Before confronting the data to model calculations, let us make some 
comments on  
how the extraction of the anisotropy ratio $R$ is
influenced by the detector cuts.
For this purpose, we used the IQMD (for both HM and SM parametrisations)
events  to compare the anisotropy ratios $R$ 
before and after passing through the apparatus acceptance.
The results are shown in Fig.~\ref{fig5}
for the PM4 multiplicity bin, in terms of the ratio of unfiltered to filtered
anisotropies, as a function of the c.m. polar angle.
As clearly evidenced by this figure, the distorsions caused
by acceptance effects appear for very small $\Theta_{\rm c.m.}$ and for 
$\Theta_{\rm c.m.} > 65^{\circ}$.  
The values of $R$ at small (large) $\Theta_{\rm c.m.}$ angles are 
mainly affected by the $\Theta_{\rm lab} = 1.2^{\circ}$ 
($\Theta_{\rm lab} = 30^{\circ}$) cut.
Within $\simeq 4^{\circ}$ and $\simeq 65^{\circ}$,
the ratio $R$ is only very little
affected (less than $5\%$ as one can see from Fig.~\ref{fig5})
by the detector apparatus. It is worth to note that the 
effect is very similar for both HM and SM calculations.
Similar observations were also made for the H parametrisation. 
In the following, the discussion will be restricted to the 
region $4^{\circ} < \Theta_{\rm c.m.} < 65^{\circ}$
where a realistic confrontation of data to model predictions
can be done with a high level of confidence, independently on how
well the experimental filter simulates the different apparatus biases.  

Using the same IQMD simulations (with a HM parametrisation),
we have also investigated~\cite{cro 96} the reliability of
the method used in the previous section to correct the experimental
anisotropy ratios for fluctuations of the reconstructed reaction plane.
The validity of this method depends on how well
the $<\cos(\Delta\Phi_{\rm R})>$ and $<\cos^2(\Delta\Phi_{\rm R})>$
quantities are estimated~\cite{dem 90}.
The latter have been calculated according to the method used for the data,
{\em i.e.} by subdividing randomly each event into two and calculating 
the half difference in azimuth ($\Delta\Phi_{\rm R}$) between the reaction
planes extracted from the two sub-events.
They are compared in Tab.~\ref{tab3} to the values obtained using the difference
in azimuth ($\Phi_{\rm R} - \Phi_{\rm true}$) between the reconstructed reaction
plane and the true one (known in the model). As one can see
the two procedures lead to very similar results,
except for the PM2 and PM3 multiplicity bins where
differences of about $5\%$ are found. This is due to the
relatively small number of detected particles per event in such
rather peripheral collisions.
We have examined the influence of this effect on the extraction
of the anisotropy ratio.
We found that even in the least favourable case PM2, the value of the 
anisotropy ratio ($R=$3.34) 
obtained within $\simeq 4^{\circ}$ and $\simeq 65^{\circ}$ by
applying the procedure used for the data to the IQMD simulated events,
remains close to the one ($R$=3.09) extracted from
the azimuthal distributions relative to the true reaction plane.

\subsection{Data versus IQMD predictions}

Tab.~\ref{tab4} shows the mean multiplicities 
of different charged ejectiles in the PM4 bin for the data 
and for IQMD filtered calculations using a HM parametrisation.
As one can see the model underpredicts by about a factor of 2.6 the 
experimental IMF multiplicities. The effect is even more important
for $Z=2$ particles (a factor of 4.3).
It is found to be most pronounced for particles emerging from the 
participant region of the collision~\cite{rei 96,ram 96,cro 96}.
Similar deviations are observed for both SM and H parametrisations.
This failure of the QMD model in reproducing the fragment multiplicities
was also reported in other works~\cite{tsa 93,gos 95}.
It might be related~\cite{gos 95} to an artificial
equilibration of the excited nuclear matter which in reality breaks up into 
several fragments.
It is also known that the clusterisation probability depends strongly on 
the gaussian wave packets width $L$ used in the model~\cite{har 96}.
Therefore, in order to perform a meaningful comparison between data and 
model predictions
({\em i.e.} with observables independent on the clusterisation
process), 
one needs to reconstruct for both data and theoretical events
coalescence invariant quantities.
This is done, as described in section III, by including in the calculations
of the anisotropy ratio all detected particles each being weighted by 
its charge.
According to this procedure, the contribution of a given nucleon to the 
resulting observable is assumed to be 
independent of the fact that this nucleon is detected
free or bound into a cluster.
It has been shown in a recent work~\cite{wan 95}
that the coalescence process
accounts for many of the observed features of the phase-space
densities of light fragments emitted from the
participant region, 
and in particular the increase in sideward flow
with fragment mass. 
All the anisotropy ratios $R$ reported in what follows are presented
in terms of such coalescence invariant quantities.

\vspace{0.5cm}

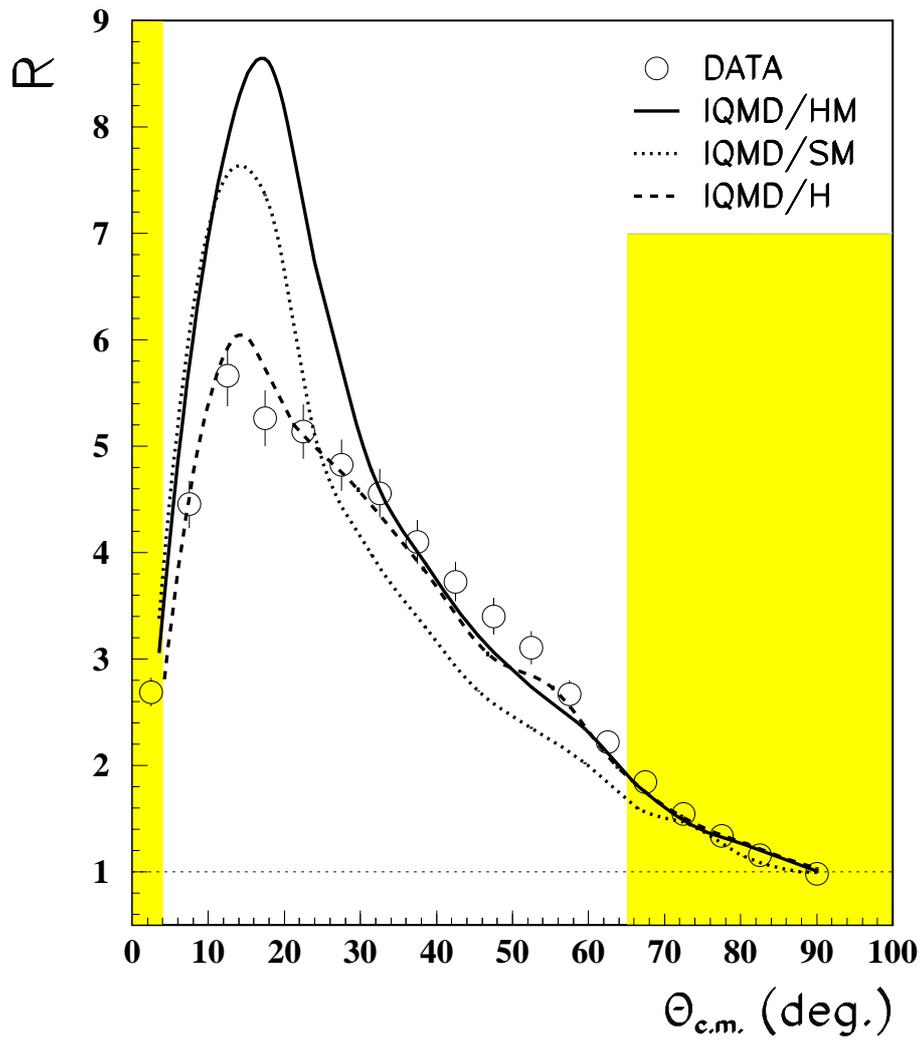
\begin{figure}
\caption{In-plane anisotropy ratio $R$
as a function of the
c.m. polar angle for semi-central (PM4) collisions.
Data (circles) are compared to the predictions of the IQMD model
with the parametrisations HM (solid curves), SM (dotted curves)
and H (dashed curves).
Theoretical values are calculated with respect to the orientation of the
true reaction plane after application of the FOPI detector filter.
Experimental values
are corrected for the unaccuracy on the determination of the reaction
plane (see text).
The shaded zone corresponds to the polar c.m. angular domain where the data
suffer significant distorsions due to apparatus effects.
Error bars on data points correspond to systematical uncertainties estimated
to $5\%$.
Statistical errors on theoretical points are lower than $2\%$.}
\label{fig6}
\end{figure}

Figure~\ref{fig6} shows the ratio $R$ as a function of $\Theta_{\rm c.m.}$ for 
semi-central 
events belonging to the PM4 multiplicity bin.
Data are compared to the predictions of the IQMD model 
for the three parametrisations HM, SM and H.
Data points are plotted with the overall systematic errors 
which have been estimated to $5\%$.
As can be seen, the general trends observed in the data are qualitatively 
well predicted by the model irrespective of the parametrisation used in the
calculations.
Indeed, theoretical $R$ values reach a maximum around 
$\Theta_{\rm c.m.} =15^{\circ}$, mainly caused by the deflection of 
the projectile remnants,
and tend toward 1 when $\Theta_{\rm c.m.}$ approaches $90^{\circ}$.
Now looking more quantitatively at the results, the following 
comments can be made.
First, around $\Theta_{\rm c.m.}=15^{\circ}$
the experimental anisotropy ratio is better reproduced 
by a hard EoS without MDI (H). 
Both SM and HM parametrisations, including MDI effects, overestimate the 
data by about $40\%$ and $60\%$, respectively.
This is disturbing as the MDI should offer a more realistic
description of the reaction since 
the proton-nucleus scattering data call for a momentum 
dependent interaction.
Thus, the previous observation would suggest that the MDI is not properly 
treated in the model.
Nevertheless it is important to point out that other parameters, 
like the in-medium ${\sigma}_{\rm nn}$
or the ones related 
to the initialisation procedure,
could also influence the calculations~\cite{har 96}.
More detailled simulations performed with different sets of parameters
should provide further information about this observation which remains
difficult to be clearly interpreted at the present level of investigation.
In particular, additionnal calculations made with a soft EoS without MDI 
would be helpfull to draw conclusions about the effect of 
the MDI with respect to the stiffness of the EoS.
It should be also noted that the use of coalescence invariant 
observables might be inappropriate at these small angles as the 
reaction mechanism behind the break-up and the evaporation of spectators 
is totally different from the freeze-out of a hot nucleon
gas~\cite{aic 91,ber 88,sto 86}.  

On the other hand, the in-plane anisotropy carried by particles emitted
from the hot and dense 
central region of the collision
({\em i.e.} at large $\Theta_{\rm c.m.}$ angles) shows a clear sensitivity 
to the stiffness of the EoS and almost no dependence to the MDI.
Data are better reproduced by both hard EoS calculations (H and HM).
The SM parametrisation leads to anisotropy ratios lower by about
$15\%$ than those predicted by the H and HM versions.
This observation agrees qualitatively with the results of previous analysis
done with other kind of in-plane flow observables
~\cite{ramil 95,wie 93,cro 96,don 96}.

Note that, by performing a similar analysis for the lower incident 
energies (100, 150, 250 and 400$A$\,MeV), we have obtained comparable 
trends but lower sensitivities, as expected in ref.~\cite{aic 87,aic 91}.

\vspace{0.5cm}

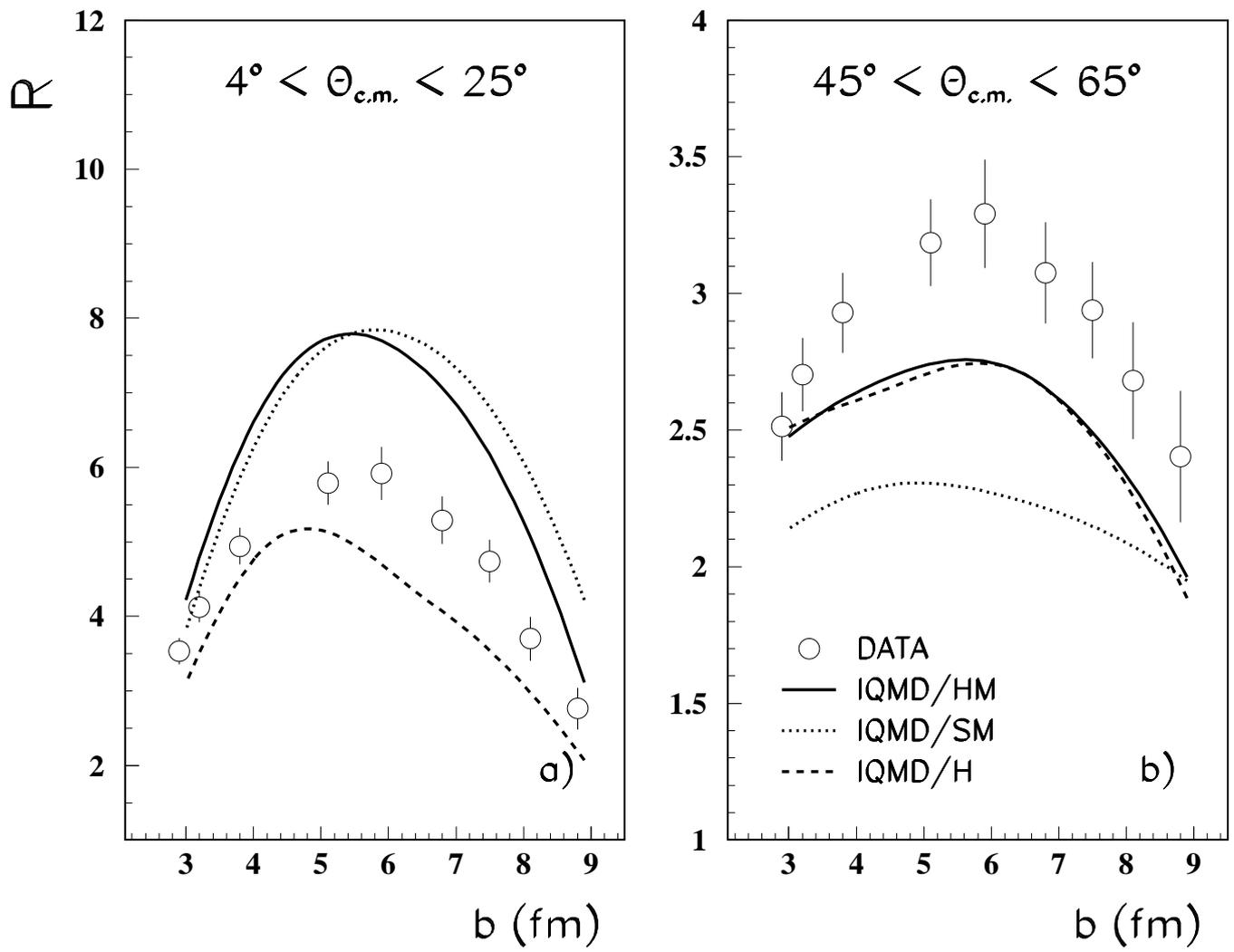
\begin{figure}
\caption{In-plane anisotropy ratio $R$
extracted in the c.m polar angular domains
$4^{\circ} < \Theta_{\rm c.m.} < 25^{\circ}$ (a) and
$45^{\circ} < \Theta_{\rm c.m.} < 65^{\circ}$ (b)
as a function of the impact parameter ($b$).
Data are compared to the predictions of filtered IQMD model calculations.
Error bars on data points correspond to systematical uncertainties estimated
between $5\%$ and $10\%$ depending on the collision centrality. 
The impact parameter is deduced from IQMD calculations using a HM 
parametrisation.
Statistical errors on theoretical points are lower than $2\%$.}
\label{fig7}
\end{figure}

Figure~\ref{fig7} shows the ratio $R$ as a function of the impact parameter.
Experimental events were sorted here into 9 multiplicity bins 
(each 5 multiplicity units wide) 
from PMUL = 20 to 65.
Theoretical $R$ values were extracted for the four large multiplicity bins 
PM2 to PM5 because of the low statistics. 
The impact parameter $b$ for the experimental data 
was determined with the IQMD model using the HM 
parametrisation (see Tab.~\ref{tab1}).
The comparison between data and model predictions is shown for 
two $\Theta_{\rm c.m.}$ regions : $4^{\circ} < \Theta_{\rm c.m.} < 25^{\circ}$ 
(Fig.~\ref{fig7}.a) and $45^{\circ} < \Theta_{\rm c.m.} < 65^{\circ}$ 
(Fig.~\ref{fig7}.b).
The latter regions were taken from 
Fig.~\ref{fig6} which shows a sensitivity of $R$ to 
the MDI and to the stiffness of the nuclear EoS for small and large 
$\Theta_{\rm c.m.}$, respectively.
In the low polar angular domain (Fig.~\ref{fig7}.a), 
the H parametrisation underestimates the experimental anisotropies, 
in particular at impact parameters above $5fm$;
while both calculations including MDI (HM and SM)
overpredict the data significantly. 
At large polar angles (Fig.~\ref{fig7}.b), all three parametrisations
underestimate the experimental results. This disagreement
is, however, less pronounced with H and HM calculations.

The comparisons in Fig.~\ref{fig6} and Fig.~\ref{fig7}
show that the general trends observed
in the data are qualitatively predicted by the model
irrespective of the parametrisation used in the calculations,
but, on a more quantitative basis none of these three model versions
(H, HM and SM)
can consistently reproduce all of the experimental results.
To understand these discrepancies, one needs more detailed
theoretical investigations including explorations of the
influence of other physical ingredients such as the in-medium
nucleon-nucleon cross section and the sensitivity of the
theoretical predictions to the input parameters
(such as the gaussian width) used for the initialisation.

As illustrated in Fig.~\ref{fig7}, 
the present data can potentially serve as
a testing ground for the theory and should allow for
disentangling compressional effects and those related to the
momentum dependence of the nuclear force.
Indeed in peripheral collisions, the anisotropy observable $R$ exhibits
an almost pure sensitivity to MDI at forward angles
(Fig.~\ref{fig7}.a) where spectator matter is predominant; while
at large ${\Theta}_{cm}$ angles ({\em i.e.} participant region) $R$ is purely
sensitive to the stiffness of the nuclear equation of state
(Fig.~\ref{fig7}.b). 
This can be understood as follows. 
For central collisions, the overlap between nuclei produces a highly dense 
nuclear matter volume where the escaping particles are strongly representative 
of the degree of compression reached during the reaction.
In addition, the high number of binary collisions between nucleons
lowers their initial relative momentum and consequently 
the repulsive effect of the MDI.
This is best seen by looking at the ratio $R$ at the more central 
collisions for large c.m. polar angles (Fig.~\ref{fig7}.b), {\em i.e.} 
by avoiding the contribution of the projectile remnants which remain 
influenced by the MDI since they feel a relative small compression 
as compared to the participant nuclear matter.
With increasing impact parameters the overlap between nuclei
({\em i.e.} compressional effects) and the number of nucleon-nucleon 
collisions decrease.
Hence, in contrast to central collisions, the ejectiles feel a higher dependence
to the MDI and a smaller dependence to the stiffness of the EoS.
In this regard, it is worth noticing that we observe the most pronounced
sensitivity to the MDI by looking at the $R$ ratio for small $\Theta_{\rm c.m.}$
angles in peripheral collisions (Fig.~\ref{fig7}.a) 
where particles are almost exclusively deflected by the initial contact 
of the colliding nuclei since compressional effects are quite low.

\vspace{0.5cm}

Although a definitive statement about the incompressibility of the nuclear
EoS remains still elusive  
at the present stage, such comparisons
have, nevertheless, the merit to illustrate the
possibilities opened with the availability
of the highly accurate data measured in the present experiment as
stringent test for microscopic transport models, and in particular,
the ability to unravel compressional and MDI effects.

\section{Summary and conclusion}

In this article, we reported the first detailled measurements of
the in-plane azimuthal anisotropy
ratio $R$ of light particles and IMF emitted in Au+Au collisions
at an incident energy of 600$A$\,MeV.
Corrections for the finite resolution of the reaction plane reconstruction
were applied.
The study of the ratio $R$ was carried-out by scanning the phase space 
in terms of 
c.m. polar angle portions. This was found to be 
better adapted 
to the present detector acceptance as compared to rapidity selections.
The $\Theta_{\rm c.m.}$ dependence of the anisotropy ratio $R$ exhibits a 
well defined maximum,
located at forward polar angles ($\Theta_{\rm c.m.} \simeq 15^{\circ}$), 
which reflects mainly the 
bounce-off of the spectator remnants.
We found that the larger the size of the detected particle, 
the stronger the magnitude of the azimuthal anisotropy 
and, therefore, the 
momentum space alignment with respect to the reaction plane.

Data were compared to the predictions of the IQMD model using three
different parametrisations of the nuclear EoS.
The comparison was performed in terms of the ratio $R$ determined
as a coalescence invariant quantity, in order to avoid the problem 
of the lack of clusterisation in the IQMD model.
It has been restricted to the c.m. polar angle domain 
$4^{\circ} < \Theta_{\rm c.m.} < 65^{\circ}$ where detector acceptance effects
on $R$ were estimated to be less than $5\%$.
We found that the general trends observed
in the data are qualitatively predicted by the model
irrespective of the parametrisation used in the calculations;
but, on a more quantitative basis, none of these three model
parametrisations (H, HM and SM)
can consistently reproduce all of the experimental results.
We show that, although a definitive statement in favour of
one of the three parametrisations cannot be made at present,
such comparisons
have, nevertheless, the merit to illustrate the
possibilities opened with the availability
of the highly accurate data measured in the present experiment,
over a broad range of impact parameters,
as providing a
stringent test for microscopic transport models.
In particular, the ability to unravel compressional and MDI effects
was clearly demonstrated : in the spectator region,
the anisotropy observable $R$ exhibits
an almost pure sensitivity to MDI, while
in the participant region $R$ is purely
sensitive to the stiffness of the nuclear equation of state.

Further detailed comparisons to microscopic transport model predictions, 
including explorations of in-medium effects on the nucleon-nucleon 
cross section,
should help to provide deeper insight into the properties of 
hot and dense nuclear matter and the underlying equation of state.

\section*{Acknowledgement}

This work was supported in part by the French-German agreement between
GSI and IN2P3/CEA and by the PROCOPE-Program of the DAAD.


\begin{table}
\caption{Accuracy on the determination of the
reaction plane for events measured under the PM2 to PM5
multiplicity conditions.
$\sigma(\Delta\Phi)$ is extracted from a gaussian fit to the $\Delta\Phi$
distributions (see text).
The limits of the different event multiplicity classes are given 
for both data (PMUL) and IQMD (PMUL$_{\mbox{\small{IQMD}}}$).
For each event class, the mean impact parameter ($<b>$) and
the corresponding r.m.s. deviation extracted
from the IQMD model
are given.}
\vspace{0.5cm}
\begin{tabular}{ccccc} 
Multiplicity bin & PM2 & PM3 & PM4 & PM5  \\ \hline
PMUL & 16-31 & 31-46 & 46-62 & $\ge$ 62 \\ \hline
PMUL$_{\mbox{\small{IQMD}}}$ & 19-38 & 38-57 & 57-76 & $\ge$ 76 \\ \hline
$ <b> (fm) $ & $8.9\pm 0.9$ & $6.7\pm 1.0$ & $4.0\pm 1.4$ & 
$3.0\pm 1.1$ \\ \hline
$ \sigma(\Delta\Phi)\; (deg.)$ & 34.5 & 23.6 & 24.7 & 30.0 \\
\end{tabular}
\label{tab1}
\end{table}

\vspace{1.cm}

\begin{table}
\caption{In-plane anisotropy ratio $R$ before and after correction
for fluctuations of the reconstructed reaction planes.
The data are shown for the $40^{\circ} < \Theta_{\rm c.m.} < 50^{\circ}$ cut.
Errors correspond to systematical uncertaincies. Statistical errors 
are lower than 1$\%$.}
\vspace{0.5cm}
\begin{tabular}{ccccc} 
Multiplicity bin & PM2 & PM3 & PM4 & PM5  \\ \hline
$R$ not corrected & $2.32\pm 0.23$ & $3.36\pm 0.24$ & $3.02\pm 0.15$ 
& $2.51\pm 0.13$ \\ \hline
$R$ corrected     & $2.86\pm 0.29$ & $3.94\pm 0.28$ & $3.55\pm 0.18$ 
& $2.96\pm 0.15$ \\
\end{tabular}
\label{tab2}
\end{table}

\vspace{1.cm}

\begin{table}
\caption{Values of $<\cos ({\Phi _{\rm R} - \Phi _{\rm true}})>$,
$<\cos {\Delta \Phi _{\rm R}}> $, 
$<\cos^2 ({\Phi _{\rm R} - \Phi _{\rm true}})> $ and 
$<\cos^2 {\Delta \Phi _{\rm R}}> $ obtained from filtered IQMD calculations.
See text for more details.}
\vspace{0.5cm}
\begin{tabular}{ccccc}
Multiplicity bin & PM2 & PM3 & PM4 & PM5 \\ \hline
$<\cos ({\Phi_{\rm R} - \Phi_{\rm true}})> $ &
$ 0.93 $ & $ 0.97 $ & $ 0.92 $ & $ 0.88 $   \\ \hline
$<\cos {\Delta \Phi _{\rm R}}>    $ &
$ 0.87 $ & $ 0.94 $ & $ 0.92 $ & $ 0.88 $   \\ \hline
$<\cos^2 ({\Phi _{\rm R} - \Phi _{\rm true}})> $ &
$ 0.90 $ & $ 0.94 $ & $ 0.89 $ & $ 0.85 $   \\ \hline
$<\cos^2 {\Delta \Phi _{\rm R}}> $ &
$ 0.85 $ & $ 0.90 $ & $ 0.87 $ & $ 0.81 $   \\
\end{tabular}
\label{tab3}
\end{table}

\vspace{1.cm}

\begin{table}
\caption{Mean multiplicities of different types of ejectiles
emitted in the forward c.m. hemisphere for PM4 events class.
Data are compared to different
filtered IQMD model simulations.
Errors represent statistical uncertaincies
including the r.m.s deviations of the distributions.}
\vspace{0.5cm}
\begin{tabular}{ccccc} 
$ $ & $Z=1$ & $Z=2$ & $Z=3$ & $3 \le Z \le 12$ \\ \hline
DATA &$44.03\pm 0.05$&$10.06\pm 0.02$&$1.72\pm 0.01$&$2.61\pm 0.01$ \\ \hline
IQMD/HM &$61.89\pm 0.27$&$2.35\pm 0.02$&$0.45\pm 0.01$&$0.99\pm 0.01$  \\
\end{tabular}
\label{tab4}
\end{table}


\end{document}